\documentstyle[12pt]{article}

\makeatletter

\@addtoreset{equation}{section}
\makeatother
\newcommand{\tselea}[1]{\label{#1}}
\newcommand{\tseleq}[1]{\label{#1}}

\newcommand{\tseref}[1]{\ref{#1}} 
\newcommand{\tsecite}[1]{\cite{#1}} 

\newcommand{\tsebibitem}[1]{\bibitem{#1}} 

\begin{document}

\typeout{--- Title page start ---}

\renewcommand{\thefootnote}{\fnsymbol{footnote}}

\begin{flushright} Imperial/TP/95-96/18 \\ 
{\tt hep-ph/9601268} \\ 
15th January 1996
\end{flushright}
\vskip 12pt

\begin{center} {\large\bf Wick's Theorem at Finite
Temperature\footnote{Available in {\LaTeX} format
through anonymous ftp from
{\tt ftp://euclid.tp.ph.ic.ac.uk/papers/95-6\_18.tex},
or on WWW in {\LaTeX} and postscript formats at
{\tt http://euclid.tp.ph.ic.ac.uk/Papers/index.html}
} }\\
\vskip 1.2cm {\large T.S. Evans\footnote{E-mail: {\tt T.Evans@ic.ac.uk},
WWW: {\tt http://euclid.tp.ph.ic.ac.uk/$\sim$time}}, D.A.
Steer\footnote{E-mail: {\tt D.Steer@ic.ac.uk}}}\\ Blackett Laboratory,
Imperial College, Prince Consort Road,\\ London SW7 2BZ  U.K.\\
\end{center}

\vskip 1cm
\begin{center} {\large\bf Abstract}
\end{center}

We consider Wick's Theorem for finite temperature and finite volume
systems.  Working at an operator level with a path ordered approach, we
 show that contrary to claims in the literature, expectation values of
normal ordered products can be chosen to be zero and that results
obtained are independent of volume.  Thus the path integral and operator
approaches to finite temperature and finite volume quantum field theories
are indeed seen to be identical.  The conditions under which normal
ordered products have simple symmetry properties are also considered.

\vskip 1cm

\renewcommand{\thefootnote}{\arabic{footnote}}
\setcounter{footnote}{0}

\typeout{--- Main Text Start ---}

\section{Introduction}

Early developments of thermal field theory were based on a path
ordered operator approach \tsecite{Ma}, and Wick's theorem played a crucial role in
the development of its perturbation theory \tsecite{Ma,Wi}.  More recently,
other approaches to the subject have been developed -- in
particular, path ordered methods have been extended, often by using
a path integral approach 
\tsecite{LvW,vW,Raybook,Mi}, and Thermo Field Dynamics
(TFD) in which a doubling of the states of
the system is introduced \tsecite{LvW,vW,Raybook,TFD,SU,He}.  It is
important to understand to what degree these various approaches are
equivalent \tsecite{LvW,vW}.  

However, a closer inspection of the original literature and subsequent
standard texts turns out to be rather confusing regarding the issue
of Wick's theorem and the expectation value of normal ordered
products.  In particular there are claims that there are finite
volume corrections to the standard thermal perturbation theory.  If
this was true then in some important cases, such as in the study of
the quark-gluon plasma formed in relativistic heavy ion collision
experiments, there would be important new corrections to some
existing calculations.  In this paper we concentrate on path-ordered
operator approaches and we clarify the confusion found in the
literature.  These results have a direct translation into a path
integral language which we consider in the penultimate section.  In terms
of TFD the principles are similar but the implementation is rather
different.  However, the case of TFD is considerably easier because
of its similarity with zero temperature field theory and the
literature reflects this.  We will consider TFD in our conclusions.

Central to our discussion is Wick's theorem which we refer to as an
operator identity rather than a relation between expectation values.
It relates time ordered products of fields to normal ordered products and
contractions.  At zero temperature, the use of Wick's theorem
revolves around the fact that normal
ordered products are defined such that annihilation operators are always
placed to the right of creation operators.  This ensures that vacuum
expectation values of normal ordered products, $N [\ldots ]$, vanish:
$\langle 0| N [a a^\dagger]|0 \rangle= 0$.  However with the ensemble
averages used in thermal field theory, products of fields normal
ordered in the same way are no longer zero: $ \sum_n e^{-n \beta \omega}
\langle n| a^\dagger a |n
\rangle \neq 0$.  Since in calculations one considers Green functions,
expectation values
of time-ordered functions, it is useful but not crucial to show that
thermal expectation values of normal ordered products are zero:  all
that is actually required is a relation between Green functions of
different orders.  It is this `expectation value version' of Wick's
theorem which is proved by Thouless \tsecite{Th} for real
relativistic scalar fields.  Using rather different techniques, Gaudin
\tsecite{Ga} proved the same `expectation value version' in a rather more
transparent way using the cyclicity of the thermal trace.
This is the approach also used in
\tsecite{Mi,FW,NO}; it avoids the need to discuss normal ordering at
all.  Volume does not enter into the argument either.  In more modern
texts \tsecite{LvW,Raybook,Ka} which rely on path integral methods, the
expectation value version of Wick's theorem is inherent in the notation
-- it is essentially encoded as Schwinger-Dyson equations for the free
field case.

We now return to the issue of normal ordering which
was not properly dealt with in the
original paper of Matsubara.  The area has been studied in subsequent
literature, and seems to be a source of confusion.  By using a
definition of normal ordering different from that of
standard field theory, Thouless \tsecite{Th} proves that thermal expectation
values of normal
ordered products of non-relativistic fermions can be set to be zero.
Fetter and Walecka \tsecite{FW} use
Gaudin's argument and claim that ``Unfortunately, $\ldots$ the ensemble
average of normal ordered products is only zero at zero temperature''
contrary to \tsecite{Th}.  Gross et al.\ \tsecite{GRH} use the same
approach.

The situation is further confused by various authors'
considerations of finite volume corrections.   Matsubara \tsecite{Ma}
suggests that the expectation values of normal ordered terms may be
zero only in the infinite volume limit.  This argument is given in
greater detail in Abrikosov et al.\
\tsecite{AGD} and Lifschitz and Pitaevskii
\tsecite{LP}.  If this was true, there could be important corrections to
standard perturbation theory, coming from the usually neglected normal
ordered terms, when working with finite volumes systems.  As the
quark-gluon plasma formed in relativistic heavy ion collisions fills a
comparatively small volume, such finite volume corrections might be of
great importance in practical calculations.  Yet where are such volume
terms hiding in the derivations using the properties of the thermal
trace \tsecite{Mi,Ga,FW,NO} and in path integral derivations
\tsecite{LvW,Raybook,Ka}?

The finite volume argument is worth repeating here for reference later
on.  For simplicity, consider a non-interacting equilibrium system of
relativistic neutral bosons at temperature $T$ lying in a volume $V$.
Let
$\widehat{\psi}(x)$ be the hermitian field operators in the Heisenberg
picture.  We wish to calculate Green functions, and as a specific
example consider the four-point thermal Wightman function
\begin{equation} {\cal{W}}_{1,2,3,4} = \ll \widehat{\psi}(x_1)
\widehat{\psi}(x_2) \widehat{\psi}(x_3) \widehat{\psi}(x_4) \gg.
\tseleq{therm1}
\end{equation}
The angular brackets denote a thermal expectation value
(which is defined in the next section), and $x_i$ are different
space-time points. In momentum space,
${\cal{W}}_{1,2,3,4}$ contains terms of the form
\begin{equation}
\frac{1}{V^2} \sum_{{\bf k}_1} \sum_{{\bf k}_2} \sum_{{\bf k}_3}
\sum_{{\bf k}_4} \ll  \hat{a}^{\dagger}_{{\bf k}_1}
\hat{a}^{\dagger}_{{\bf k}_2}  \hat{a}_{{\bf k}_3} \hat{a}_{{\bf
k}_4}\gg  {\rm exp}({\cal{Y}}),
\tseleq{conf1}
\end{equation}
where
$\hat{a}^{\dagger}_{\bf k}$ and $\hat{a}_{\bf k}$ are creation and 
annihilation 
operators, the ${\bf k}_i$'s are discrete momenta, and
$\cal{Y}$ is a sum of scalar products of 4 momenta and space-time
coordinates.  In the literature it is claimed that there are three
separate contributions to (\tseref{conf1}) \tsecite{Ma, AGD, LP}.  The
first 2 contributions arise when the momenta are equal in pairs, that is
when ${\bf k}_1= {\bf k}_3$, ${\bf k}_2={\bf k}_4$ and ${\bf k}_1= {\bf
k}_4$, ${\bf k}_2={\bf k}_3$.  For example, in the former case
(\tseref{conf1}) becomes
\begin{equation}
\frac{1}{V^2} \sum_{{\bf k_1}} \sum_{{\bf k}_2} \ll
\hat{a}^{\dagger}_{{\bf k}_1}
\hat{a}^{\dagger}_{{\bf k}_2}  \hat{a}_{{\bf k}_1} \hat{a}_{{\bf
k}_2}\gg  {\rm exp}({\cal{Y}}).
\tseleq{conf2}
\end{equation}
The final contribution arises when all the momenta are
equal:
\begin{equation}
\frac{1}{V^2} \sum_{\bf k} \ll \hat{a}^{\dagger}_{\bf k}
\hat{a}^{\dagger}_{\bf k}  \hat{a}_{\bf k} \hat{a}_{\bf k} \gg  {\rm
exp}({\cal{Y}}).
\tseleq{conf3}
\end{equation}
According to some standard texts (for example
\tsecite{ AGD,LP}), ``expression (\tseref{conf2}) has a special feature
which distinguishes it from (\tseref{conf3}), {\it i.e.}\ the number of
factors in $1/V$ is the same as the number of summations, whereas in
(\tseref{conf3}) this number of factors is larger''.  Thus
(\tseref{conf3}) has ``an extra factor of
$1/V$ and will vanish as $V \rightarrow \infty$''.  These texts conclude
that, once the infinite volume limit has been taken, ``we can actually
average over pairs of operators''.  The implication is that only in this
limit does the four-point Green function (or in general an $n$-point
one) split into a sum of products of two-point Green functions.

In this paper we prove that the above argument is incorrect and that any
$n$-point Green function can be decomposed exactly into two-point Green
functions, {\em whatever} $V$.  We also prove explicitly that thermal
expectation values of normal ordered products of {\em any} type of field
may always set to zero, contrary to claims elsewhere \tsecite{FW}.  The
symmetry properties of normal ordered products are considered.

The paper is organised as follows.  In section 2 we set up our notation
and definitions.  This is essential as we will be discussing different
types of normal ordering and we wish to track volume factors very
carefully.  In section 3 we split fields into two parts at an operator
level and define normal ordering and contractions appropriately.  We
choose a split which guarantees that the thermal expectation value of
normal ordered products of any two field operators is always zero and
hence that the normal ordered products have simple symmetry properties.
We then prove in section 4 that this is sufficient to ensure that the
thermal expectation value of any four-point normal ordered product also
vanishes, independent of volume.  With Wick's theorem, this  then shows
directly that the four-point Green function decomposes exactly into
two-point Green functions.  In section 5 we generalise this result from
four-point to $n$-point Green functions by using more powerful methods.
One is the usual path integral approach to thermal field theory.  The
other is based on the cyclic property of the trace which defines the
thermal expectation value \tsecite{Ga,FW}.  We use these, with Wick's
theorem, to show that the thermal expectation value of all $n$-point
normal ordered products vanish.  We also note the symmetry of the
$n$-point normal ordered products.

\section{Definitions and Notation}

For simplicity we consider only free field theories and work in the
Heisenberg picture.  Our results are completely equivalent to working
with an interacting theory in the interaction picture.  There the
operators evolve according to the free Hamiltonian and the interactions
are encoded in the $S$ matrix.  When expanded, the $S$ matrix leads one
to consider time-ordered products which are of the same form as those
considered here.  At finite temperature, one has to be rather more
careful about the use of an $S$ matrix, but one can still follow a
similar algorithm so that our results can be applied to practical
calculations at finite temperature.

Field operators, $\widehat{\psi}(x)$, are  given in terms of particle
and anti-particle creation and annihilation operators,
$\hat{a}^{\dagger}_{\bf k}$, $\hat{a}_{\bf k}$ and
$\hat{b}^{\dagger}_{\bf k}$, $\hat{b}_{\bf k}$ respectively;
\begin{eqnarray}
\widehat{\psi}(x) =
\sum_{ \bf k} \frac {1}{\sqrt{ 2 \omega_{\bf k} V}}
\left(\hat{a}_{\bf k} e^{-ik.x} + \hat{b}^{\dagger}_{\bf k} e^{ik.x}
\right),
\tselea{psi}
\\
\widehat{\psi}^{\dagger}(x) = \sum_{\bf k} \frac {1}{\sqrt{ 2 \omega_{\bf
k} V}} \left(\hat{b}_{\bf k} e^{-ik.x} + \hat{a}^{\dagger}_{\bf k}
e^{ik.x} \right).
\tselea{psidag}
\end{eqnarray}
The metric is ${\rm diag} (+1,-1,-1,-1)$ and $\omega_{\bf
k}^2 = {\bf k}^2 + m^2$ for relativistic field theories.  For
non-relativistic ones $\omega_{\bf k} = {\bf k}^2/2 m$, and
$\hat{b}_{\bf k} = 0$.  A real relativistic scalar theory has
$\hat{a}_{\bf k} = \hat{b}_{\bf k}$.  For fermionic theories,
$\hat{a}_{\bf k}$ and $\hat{b}_{\bf k}$ implicitly carry spin indices,
and there is an implicit sum over spins as well as over momenta.
Annihilation and creation operators must also be multiplied by the
correct solutions to the Dirac equation if the theory is relativistic
but this does not effect our results.   Likewise, the analysis holds for
fields of higher spin, where additional indices have been suppressed, as
only the statistics of the field is relevant to this analysis.

The operators obey the canonical commutation relations,
\begin{eqnarray}
\left[ \hat{a}_{\bf k}, \hat{a}^{\dagger}_{\bf k'} \right]_{\sigma}  & =
& \delta_{{\bf k},{\bf k'}} \; = \; \hat{a}_{\bf k}
\hat{a}^{\dagger}_{\bf k'}  - \sigma  \hat{a}^{\dagger}_{\bf k'}
\hat{a}_{\bf k}
\tselea{comma} \\
\left[ \hat{b}_{\bf k}, \hat{b}^{\dagger}_{\bf k'} \right]_{\sigma} & =
& \delta_{{\bf k},{\bf k'}}
\tselea{commb} \\
\left[ \hat{a}_{\bf k}, \hat{b}^{\dagger}_{\bf k'} \right]_{\sigma} & =
& 0,
\tselea{commbc}
\end{eqnarray}
where $\sigma = +1$ for bosonic operators, and $\sigma =
-1$ for fermionic ones.

The free Hamiltonian is
\begin{equation}
\hat{H}_0  = \sum_{\bf k} \omega_{\bf k} \left[
\hat{N}^{a}_{\bf k} + \hat{N}^{b}_{\bf k} \right] + \sum_{\bf k} \left(
\mu_{a} \hat{N}^{a}_{\bf k} + \mu_{b} \hat{N}^{b}_{\bf k} \right) .
\tseleq{ham}
\end{equation}
The $\mu_{a,b}$ are chemical potentials for particles and
anti-particles respectively.  If we have a U(1) symmetry, as in the case
of a complex relativistic field, then
$\mu_a = -\mu_b$. For a real scalar field $\mu_a = \mu_b=0$.  The
$\hat{N}^{a,b}_{\bf k}$ are conserved number  operators whose normalised
eigenvectors are denoted by
$|n_{\bf k} \rangle$ for particles, and $|m_{\bf k} \rangle$ for
anti-particles.  They are complete and satisfy
\begin{equation}
\hat{N}^{a}_{\bf k} |n_{\bf k} \rangle = n_{\bf k}  |n_{\bf k} \rangle.
\tseleq{evectora}
\end{equation}
For systems described by Bose-Einstein statistics, $n_{\bf
k} = 0, 1, 2, \ldots$, whilst for Fermi-Dirac systems, $n_{\bf k} = 0,
1$.


Thermal expectation values are denoted by angular brackets, and are
defined by
\begin{eqnarray}
\ll \ldots \gg  & = & \frac{1}{Z_0} {\rm Tr} \left\{ e^{-\beta
\hat{H}_0 } \ldots \right\} \nonumber \\
  & = & \frac{1}{Z_0}  \sum_{ \{ n_{\bf k} \}, \{ m_{\bf k} \}}
\langle \{ n_{\bf k} \}, \{ m_{\bf k} \}|  e^{-\beta \hat{H}_0}
\ldots  |\{ n_{\bf k} \}, \{ m_{\bf k} \} \rangle,
\tselea{thermexpect}
\end{eqnarray}
where $\ldots$ means any operator,  and $\beta$ is the
inverse temperature: $\beta = 1/T$ ($k_B = 1$).  The trace has been
taken over the number operator eigenstate basis
$|\{n_{\bf k}\}, \{m_{\bf k}\} \rangle= |n_{{\bf k}_1}\rangle
\otimes  |n_{{\bf k}_2 }\rangle \otimes
\ldots |m_{{\bf k}_1}\rangle \otimes  |m_{{\bf k}_2 }\rangle
\ldots$.  $Z_0$ is the free partition function of the system:
\begin{eqnarray} Z_0  & = & {\rm Tr} \left\{ e^{-\beta
\hat{H}_0} \right\} \nonumber
\\
 & = & \left( \prod_{\bf k} Z^{a}_{\bf k} \right) \left( \prod_{\bf p}
Z^{b}_{\bf p} \right).
\tselea{partfn}
\end{eqnarray}
$Z^{a,b}_{\bf k}$ is related to the Bose-Einstein and Fermi-Dirac
distributions
\begin{equation} N^{a,b}_{\bf k} = \left( e^{\beta ( \omega_{\bf k}  +
\mu_{a,b}) } - \sigma \right) ^{-1}
\tseleq{BEFD}
\end{equation}
by
\begin{equation} Z^{a,b}_{\bf k} = \left( 1 + N_{\bf k}^{a,b}
\right)^{\sigma}.
\tseleq{partfnk}
\end{equation}

The following identities are useful for calculating thermal expectation
values of sets of creation and annihilation operators
\begin{equation}
\sum_{n_{\bf k} = 0}^{\infty} n_{\bf k} e^{-\beta (\omega_{\bf k} +
\mu_{a,b}) n_{\bf k}} = N^{a,b}_{\bf k} \left( 1 + N^{a,b}_{\bf k}
\right),
\tseleq{rel1}
\end{equation}
\begin{equation}
\sum_{n_{\bf k} = 0}^{\infty} (n_{\bf k})^2 e^{-\beta (\omega_{\bf k} +
\mu_{a,b}) n_{\bf k}} = N^{a,b}_{\bf k} \left( 1 + N^{a,b}_{\bf k}
\right) \left( 1 + 2 N^{a,b}_{\bf k} \right) .
\tseleq{rel2}
\end{equation} Thus for fields of either statistic,
\begin{eqnarray}
\ll \hat{a}^{\dagger}_{\bf k} \hat{a}_{\bf p} \gg  & = & N^{a}_{\bf k}
\delta_{{\bf k},{\bf p}}
\tselea{b1}
\\
\ll \hat{a}_{\bf k} \hat{a}^{\dagger}_{\bf p} \gg  & = & \left( 1 +
\sigma N^{a}_{\bf k} \right) \delta_{{\bf k},{\bf p}}
\tselea{b2}
\end{eqnarray}

We now split the field operators arbitrarily into two parts
\begin{equation}
\widehat{\psi}(x) = \widehat{\psi}^{+}(x) + \widehat{\psi}^{-}(x).
\tseleq{decomp}
\end{equation}
We call the parts ``positive'' and ``negative'' even
though in general the split is {\em not} into positive and negative
energy waves.   In order to avoid possible confusion between $\dagger$
and $+$ superscripts, the hermitian conjugate of an operator is denoted
by $\bar{\psi}  = {\psi}^{\dagger} $.   We do not assume that the
operator and its hermitian conjugate are split in the same way.  That is,
we do not assume  that $\left( [\bar{\psi}]^{+}  \right)^{\dagger} =
{\psi}^{\pm} $.  This means care must be taken with the notation; we
always have $\bar{\psi}^{+}   = [\psi^\dagger]^{+} $ rather than it being
equal to $[{\psi}^{\pm} ]^\dagger$.

To allow for products of more than one type of field operator, we
now use the notation
\begin{equation}
\psi_i = \widehat{\psi}_i(x_i) .
\end{equation}
Of course we will often be looking at products where some
or all of the operators $\widehat{\psi}_i(x)$ are equal or related by
hermitian conjugation.

Normal ordering is defined in terms of the arbitrary split (\tseref{decomp}),
so
generalising the traditional $T=0$ definition which is expressed in terms
of annihilation and creation operators.  We {\em always} define
normal ordering to strictly mean that all $+$ fields are placed to the
right of $-$ fields; otherwise the order of the fields is left unchanged.
(So it differs from that used for fermions in \tsecite{Th}.)  For
example,
\begin{equation} N \left[ \psi_1 \psi_2 \right] := N_{1,2} =
\psi_{1}^{+} \psi_{2}^{+} +
\psi_{1}^{-} \psi_{2}^{+} + \sigma \psi_{2}^{-} \psi_{1}^{+} +
\psi_{1}^{-} \psi_{2}^{-}.
\tseleq{N2}
\end{equation}
Here $\psi_{1}$ and $\psi_{2}$ are {\em any} two fields
with different space-time arguments $(t_1,{\bf x}_1)$ and $(t_2,{\bf
x}_2)$ respectively (which for convenience we usually suppress).   The
fields may be related, e.g.\ one often has
${\psi}_{1} = \psi(t_1,{\bf x}_1)^\dagger,
\psi_{2}=\psi(t_2,{\bf x}_2)$.   For general splits $N_{1,2}$ may
{\em not} be symmetric;  $N_{1,2} \neq \sigma N_{2,1}$.

Time ordering is defined in the usual way so for the two-point case we
have
\begin{equation}
T \left[ \psi_1 \psi_2 \right] := T_{1,2} = \psi_1
\psi_2 \theta(t_1 - t_2) + \sigma \psi_2 \psi_1 \theta(t_2 - t_1).
\tseleq{T2}
\end{equation}
It is straightforward to generalise this time ordering
to path ordering as used in some finite temperature analysis.  Then the
theta functions of time in (\tseref{T2}) become defined in terms of a
path in complex time \tsecite{LvW}.  As this does not effect the
subsequent analysis we will not complicate our notation further.


Wick's theorem is an operator identity.  For two-point functions it
relates $T_{1,2}$ to $N_{1,2}$ by defining the contraction
$D [ \psi_1 \psi_2 ] = D_{1,2}$:
\begin{eqnarray} D_{1,2} & :=  & T_{1,2} - N_{1,2}
\tselea{W2}
\\ & = & \theta(t_1 - t_2) \left[ \psi_{1}^{+}, \psi_{2}^{-} \right]
_{\sigma}
\nonumber
\\ &  & - \; \theta(t_2 - t_1) \left\{
\left[ \psi_{1}^{+}, \psi_{2}^{+} \right] _{\sigma} +
\left[ \psi_{1}^{-}, \psi_{2}^{+} \right] _{\sigma} +
\left[ \psi_{1}^{-}, \psi_{2}^{-} \right] _{\sigma} \right\}.
\tselea{W2b}
\end{eqnarray}
We will only consider splits, (\tseref{decomp}),  for
which $D_{1,2}$ is a c-number -- that is splits
linear in the annihilation and creation operators.  Note that the
contraction, and hence the normal ordered product, is only symmetric,
$D_{1,2} = \sigma D_{2,1}$, if
\begin{equation}
\left[ \psi_{1}^{+}, \psi_{2}^{+} \right] _{\sigma} +
\left[ \psi_{1}^{-}, \psi_{2}^{-} \right] _{\sigma} = 0.
\tseleq{Dsymm}
\end{equation}

To higher order, Wick's theorem can be obtained from the
generating functional
\begin{eqnarray}
\lefteqn{T \left[ \exp \{ -i \int d^{4} x \; j_i(x) \psi_i(x) \}
\right] =
N \left[ \exp \{ -i \int d^{4} x \; j_i(x) \psi_i(x) \} \right] }
\nonumber
\\
& \times &
\exp \{ - \frac{1}{2} \int d^{4} x \; d^{4} y \;
j_i(x) D [ \psi_i(x) \psi_j(y) ] j_j(y) \},
\tselea{genWick}
\end{eqnarray}
where $j_i(x)$ are sources.  Functional differentiation with
respect to these sources gives Wick's theorem, which after using the
symmetry properties of the time-ordered  product is of the form
\begin{equation}
T_{1,2, \ldots , 2n} = \sum_{\rm perm} (-1)^p \sum_{m=0}^{n}
\left[ \frac{N_{a_1,a_2,\ldots, a_{2m}}}{(2m)!} .
\frac{1}{(n-m)!}
\left( \prod_{j=m+1}^{n} \frac{1}{2}D_{a_{2j-1},a_{2j} }
\right) \right] .
\tseleq{Wngen}
\end{equation}
The sum takes $a_1, \ldots a_{2n}$ through all
permutations of $1, \ldots ,2n$, and $p$ is the number of times one
has to interchange pairs of
fermions in moving from $1, \ldots ,2n$ to $a_1, \ldots a_{2n}$
order.  The sum over permutations can be simplified (even for
non-symmetric normal ordered products and contractions) to give the
usual form of Wick's theorem.  For four bosonic fields,
(\tseref{Wngen}) gives the well known result
\begin{eqnarray}
T_{1,2,3,4}  & = & N_{1,2,3,4} + D_{1,2}N_{3,4} +
D_{1,3}N_{2,4} +D_{1,4}N_{2,3} + D_{2,3}N_{1,4} +
\nonumber
\\
& &
D_{2,4}N_{1,3} + D_{3,4}N_{1,2} + D_{1,2}D_{3,4} + D_{1,3}D_{2,4}
+ D_{1,4}D_{2,3} .
\tselea{W4}
\end{eqnarray}
To order $2n$ one has
\begin{eqnarray}
T_{1,2, \ldots , 2n}& = & N_{1,2,\ldots , 2n}
\nonumber
\\
& + &  D_{1,2} N_{3,4,\ldots , 2n} + \mbox{ all other terms  with one
contraction}
\nonumber
\\
& + & D_{1,2}D_{3,4} N_{5,6,\ldots , 2n} + \mbox{ all other terms
with two contractions}
\nonumber
\\
& \vdots &
\nonumber
\\
& + & D_{1,2}D_{3,4} \ldots D_{2n-1, 2n} + \mbox{ all other
possible pairings} .
\tselea{Wn}
\end{eqnarray}
Once again, we stress that (\tseref{genWick}) -- (\tseref{Wn}) hold whatever
the
decomposition of the fields into positive and negative parts
{\em provided the contraction is a c-number}. They hold whether or not
the contraction is symmetric.

\section{Two-point functions}

In this section we prove that it is always possible to define a split
$\widehat{\psi}(x) = \widehat{\psi}^{+}(x) + \widehat{\psi}^{-}(x)$ such
that $\ll N_{1,2} \gg = 0$.  This then ensures that $D_{1,2} = \ll
T_{1,2} \gg = \sigma D_{2,1}$.

A general linear decomposition of
${\psi}(x)$ and $\bar{\psi}(x)$ into positive and negative parts is
\begin{equation}
{\psi}^{+}(x) = \sum_{\bf k} V_{\bf k}
\left( (1 - f_{\bf k}) \hat{a}_{\bf k} e^{-ik.x} + F_{\bf k}
\hat{b}^{\dagger}_{\bf k} e^{ik.x} \right),
\tseleq{psiplus}
\end{equation}
\begin{equation}
{\psi}^{-}(x) = \sum_{\bf k} V_{\bf k}
\left( f_{\bf k} \hat{a}_{\bf k} e^{-ik.x} + (1 - F_{\bf k})
\hat{b}^{\dagger}_{\bf k} e^{ik.x} \right),
\tseleq{psiminus}
\end{equation}
\begin{equation}
\bar{\psi}^{+}(x) = \sum_{\bf k} V_{\bf k}
\left( g_{\bf k} \hat{a}^{\dagger}_{\bf k} e^{ik.x} + (1 - G_{\bf k})
\hat{b}_{\bf k} e^{-ik.x} \right),
\tseleq{psidagplus}
\end{equation}
\begin{equation}
\bar{\psi}^{-}(x) =  \sum_{\bf k} V_{\bf k}
\left( (1 - g_{\bf k}) \hat{a}^{\dagger}_{\bf k} e^{ik.x} + G_{\bf k}
\hat{b}_{\bf k} e^{-ik.x} \right) .
\tseleq{psidagminus}
\end{equation}
Here $V_{\bf k} = 1/\sqrt{ 2 \omega_{\bf k} V}$, and
$f_{\bf k}$,
$g_{\bf k}$, $G_{\bf k}$ and $F_{\bf k}$ are arbitrary complex numbers.
We now impose the two independent conditions
\begin{eqnarray}
\ll N \left[ {\psi}_1 {\psi}_2 \right] \gg & = & 0,
\tselea{C1}
\\
\ll N \left[ \psi_2 {\psi}_1  \right] \gg & = & 0 .
\tselea{C2}
\end{eqnarray}
We find that (\tseref{C1}) and
(\tseref{C2}) are satisfied if and only if
\begin{eqnarray}  f_{\bf k} g_{\bf k}  & = & - \sigma N^{a}_{\bf k},
\nonumber
\\ (1 - f_{\bf k}) (1 - g_{\bf k}) & = & \left( 1 + \sigma N^{a}_{\bf k}
\right),
\tselea{split1}
\\ F_{\bf k} G_{\bf k}  & = & - \sigma N^{b}_{\bf k}, \nonumber
\\ (1 - F_{\bf k}) (1 - G_{\bf k}) & = & \left( 1 + \sigma N^{b}_{\bf k}
\right),
\tselea{split2}
\end{eqnarray}

First consider (\tseref{split1}) and (\tseref{split2}) in the limit $T
\rightarrow 0$.  They are solved by $f_{\bf k} = g_{\bf k}= G_{\bf k}=
F_{\bf k}=0$ which is just the traditional $T=0$ split as all the
annihilation operators are in ${\psi}^{+}$ and the creation operators
in ${\psi}^{-}$.  Taking the vacuum expectation value of
(\tseref{W2}), the two-point version of Wick's theorem, we see that for
$T=0$, $D_{1,2}= \langle 0| T [ \psi_1 \psi_2] |0\rangle =
G_{1,2}(T=0)$,
the zero temperature propagator.

For $T>0$, the equations for $f_{\bf k}$ and $g_{\bf k}$, $F_{\bf k}$
and $G_{\bf k}$ respectively each have two solutions
\begin{eqnarray}
f_{\bf k} = -\sigma N_{\bf k}^a + s^a
[N_{\bf k}^a(N_{\bf k}^a+\sigma)]^{1/2}
&,&
g_{\bf k} = -\sigma N_{\bf
k}^a - s^a [N_{\bf k}^a(N_{\bf k}^a+\sigma)]^{1/2},
\tselea{split1soln}
\\
F_{\bf k} = -\sigma N_{\bf k}^b + s^b [N_{\bf k}^b(N_{\bf
k}^b+\sigma)]^{1/2}
&,&
G_{\bf k} = -\sigma N_{\bf k}^b - s^b
[N_{\bf k}^b(N_{\bf k}^b+\sigma)]^{1/2},
\tselea{split2soln}
\end{eqnarray}
where $s^{a,b}=\pm1$ gives the two solutions in each
case. Now, again, the thermal expectation value of (\tseref{W2}) gives
$D_{1,2} = \ll T [\psi_1 \psi_2] \gg = G_{1,2}(T)$, where $G_{1,2}(T)$
is the thermal propagator.

For large ${\bf k}$, we see that $f_{\bf k}, g_{\bf k}, F_{\bf k}$
and $G_{\bf k}$ all tend to zero.  Thus at high energies
the split which gives zero normal ordered products at finite
temperature is tending to be the same as the usual one used at zero
temperatures.  This is not suprising as the ultra-violet behaviour
of finite temperature field theory is well known to be the same as
zero temperature field theory.

For bosons, (\tseref{split1soln}) and (\tseref{split1soln}) give real
$f_{\bf k}, g_{\bf k}, F_{\bf k}$
and $G_{\bf k}$, but $\left( [\bar{\psi}]^{+}
\right)^{\dagger} \neq {\psi}^{\pm}$ unless $T=0$.

For fermions $f_{\bf k} = g_{\bf k}^*$ and  $|f_{\bf k} |^2 =
|g_{\bf k} |^2 = N_{\bf k}^a$,  and similarly for $F_{\bf k}$ and $G_{\bf k}$.
However, in this case, $\left( [\bar{\psi}]^{+}
\right)^{\dagger} = {\psi}^{-}$ for all $T$.

For non-relativistic fields, we just set $\hat{b}_{\bf k}=0$.  In the
fermionic case, this recaptures the result of Thouless
\tsecite{Th}.

For real relativistic scalar fields there is only one type of creation
and annihilation operator.  As $\hat{a}_{\bf k}=\hat{b}_{\bf k}$ and
$N^{a}_{\bf k} = N^{b}_{\bf k}=N_{\bf k}$, a single
Bose-Einstein distribution, one must have
$F_{\bf k} = g_{\bf k}$, $G_{\bf k} = f_{\bf k}$ in
(\tseref{psiplus})-(\tseref{psidagminus}).  This gives a general linear
decomposition of a real scalar field.  Equations (\tseref{C1}) and
(\tseref{C2}) are now equivalent and their solutions are as above.

Finally we consider the question of the symmetry of the normal
ordered products for the case of two-point functions.  On substituting
(\tseref{psiplus}) to (\tseref{psidagminus}) into (\tseref{Dsymm}) one
finds that
\begin{equation}
D_{1,2} = \sigma D_{2,1}
\Leftrightarrow N_{1,2} = \sigma N_{2,1}  \Leftrightarrow  f_{\bf k} +
g_{\bf k} = 2 f_{\bf k} g_{\bf k} \; , \; \; F_{\bf k} + G_{\bf k} = 2
F_{\bf k} G_{\bf k} .
\tseleq{symmno}
\end{equation}
This means that the split traditionally used in zero
temperature work, where
$f_{\bf k} = g_{\bf k} = F_{\bf k} = G_{\bf k} =0$, has a
symmetric two-point normal ordered product. However
(\tseref{split1soln}) and (\tseref{split2soln}) also satisfy
(\tseref{symmno}).  Thus the splits for which the thermal expectation
value of two-point normal ordered products are zero, are guaranteed to
have symmetric two-point normal ordered products.

\section{Four-point functions}

We now prove that the thermal expectation value of the four-point normal
ordered products $N_{1,2,3,4}$ vanish for the split of section 3.  Hence
we show that {\em the four-point Green function decomposes into two-point
Green functions for all $V$}.  For simplicity, we consider a real
relativistic scalar field theory.  Our results are generalised to other
theories in section 5.

Equation (\tseref{W4}) can be used to relate
$N_{1,2,3,4}$ to $N^{'}_{1,2,3,4}$, where in the primed normal ordered
product the operators have been split according to the
$T=0$ split ({\it i.e.} the one for which  $f_{\bf k} = g_{\bf k}=0$).
The thermal expectation values of these two quantities satisfy
\begin{eqnarray}
\ll N_{1,2,3,4} \gg \;
& = &
\; \ll N^{'}_{1,2,3,4} \gg - \;
\left(
D^{'}_{1,2} -  D_{1,2} \right) \left( D^{'}_{3,4} - D_{3,4}
\right)
\nonumber
\\
& - &
\left(
D^{'}_{1,3} -  D_{1,3} \right) \left( D^{'}_{2,4} - D_{2,4}
\right)
-
\left(
D^{'}_{1,4} -  D_{1,4} \right) \left( D^{'}_{2,3} - D_{2,3}.
\right)
\tselea{N4Nprimed4}
\end{eqnarray}
Using the results
of the previous section, it can be shown that
\begin{equation}
 D^{'}_{i,j} -  D_{i,j}  =
\sum_{\bf k} (V_{\bf k})^{2} N_{\bf k}
\left[ e^{-ik.(x_i - x_j)} + e^{ik.(x_i - x_j)} \right].
\tseleq{DprimedminusD}
\end{equation}
The only terms which contribute to $\ll N^{'}_{1,2,3,4}
\gg $ must contain 2 positive and 2 negative parts.  There are 6 such
terms, and one of them is
\begin{eqnarray}
& \ll \widehat{\psi}^{-}(x_1) \widehat{\psi}^{-}(x_2)
\widehat{\psi}^{+}(x_3) \widehat{\psi}^{+}(x_4)  \gg  &
\nonumber
\\
= & \sum_{{\bf k}_1, {\bf k}_2,{\bf k}_3,{\bf k}_4} V_{{\bf k}_1}
V_{{\bf k}_2} V_{{\bf k}_3} V_{{\bf k}_4}
\left( \ll \hat{a}^{\dagger}_{{\bf k}_1} \hat{a}^{\dagger}_{{\bf k}_2}
\hat{a}_{{\bf k}_3} \hat{a}_{{\bf k}_4} \gg \right)  e^{ik_1 \cdot x_1}
e^{ik_2 \cdot x_2} e^{-ik_3 \cdot x_3}  e^{-ik_4 \cdot x_4}. &
\tselea{N4explicit}
\end{eqnarray}
As in the introduction, we consider $\ll
\hat{a}^{\dagger}_{1} \hat{a}^{\dagger}_{2}
\hat{a}_{3} \hat{a}_{4} \gg$ where $\hat{a}_{i} = \hat{a}_{{\bf k}_i}$,
and separate out explicitly the three contributions:
\begin{eqnarray}
\ll \hat{a}^{\dagger}_{1} \hat{a}^{\dagger}_{2}
\hat{a}_{3} \hat{a}_{4} \gg
& = &
\delta_{1, 3} \delta_{2, 4}
\ll \hat{a}^{\dagger}_{1} \hat{a}^{\dagger}_{2}
\hat{a}_{1} \hat{a}_{2} \gg \left( 1 -
\delta_{1, 2} \right)
\nonumber
\\
 & + &
\delta_{1, 4} \delta_{2, 3}
\ll \hat{a}^{\dagger}_{1} \hat{a}^{\dagger}_{2}
\hat{a}_{2} \hat{a}_{1} \gg \left( 1 -
\delta_{1, 2} \right)
\nonumber
\\
 & + &
\delta_{1, 2} \delta_{1, 3}
\delta_{3, 4}
\ll \hat{a}^{\dagger}_{1} \hat{a}^{\dagger}_{1}
\hat{a}_{1} \hat{a}_{1} \gg.
\tselea{explicit}
\end{eqnarray}
Here $\delta_{1,2} = \delta_{{\bf k}_1,{\bf k}_2}$.
{}From section 2,
\begin{equation}
\ll \hat{a}^{\dagger}_{1} \hat{a}^{\dagger}_{2}
\hat{a}_{1} \hat{a}_{2} \gg \;  =  N_{1} N_{2} = \;
\ll \hat{a}^{\dagger}_{1} \hat{a}^{\dagger}_{2}
\hat{a}_{2} \hat{a}_{1} \gg,
\tseleq{cont12}
\end{equation}
where $N_{i} = N_{{\bf k}_i}$, and
\begin{eqnarray}
\ll \hat{a}^{\dagger}_{1} \hat{a}^{\dagger}_{1}
\hat{a}_{1} \hat{a}_{1} \gg & = &
\ll \hat{a}^{\dagger}_{1} \hat{a}_{1}
\hat{a}^{\dagger}_{1} \hat{a}_{1} \gg  - \ll \hat{a}^{\dagger}_{1}
\hat{a}_{1} \gg
\nonumber
\\
 & = &  N_{1} \left( 2 N_{1} + 1
\right) - N_{1}
\nonumber
\\
 & = &  2  \left( N_{1} \right) ^2.
\tselea{cont3}
\end{eqnarray}  Hence (\tseref{explicit}) reduces to
\begin{eqnarray}
\ll \hat{a}^{\dagger}_{1} \hat{a}^{\dagger}_{2}
\hat{a}_{3} \hat{a}_{4} \gg  & = &
\left( \delta_{1, 3} \delta_{2, 4} +
\delta_{1, 4} \delta_{2, 3} \right) N_{1}N_{2}
\nonumber
\\
 & = &
\left( \delta_{1, 3} \delta_{2, 4} +
\delta_{1, 4} \delta_{2, 3} \right)
\ll \hat{a}^{\dagger}_{1} \hat{a}_{1} \gg
\ll \hat{a}^{\dagger}_{2} \hat{a}_{2} \gg.
\tselea{thatsit}
\end{eqnarray}  Given (\tseref{thatsit}), (\tseref{N4explicit}) and
(\tseref{DprimedminusD}) it is easy to evaluate the right hand side of
(\tseref{N4Nprimed4}) and find that
\begin{equation}
\ll N_{1,2,3,4} \gg = 0 .
\tseleq{N40}
\end{equation} Thus taking the thermal expectation value of
(\tseref{W4}), the four-point Green function $G_{1,2,3,4}$ is seen to
decompose exactly into two-point Green functions $G_{1,2}$:
\begin{equation} G_{1,2,3,4} = G_{1,2}G_{3,4} + G_{1,3}G_{2,4} +
G_{1,4}G_{2,3}.
\tseleq{G4}
\end{equation}

Equation (\tseref{thatsit}) shows explicitly how contributions down by
a factor of the volume, such as (\tseref{conf3}), which is the last term
of (\tseref{explicit}), do in fact cancel.  We just have to be careful
to exclude the contribution of (\tseref{conf3}) from (\tseref{conf2}) --
hence the $(1 - \delta)$ terms in (\tseref{explicit}).
It is clear from this derivation that there are no volume corrections
to expectation value version of Wick's theorem (\tseref{G4}).

\section{$n$-point functions}

We now generalise (\tseref{G4}) to $n$-point Green functions by using
more powerful methods, namely the ``trace theorem'' and the path
integral.  Both of these methods may be applied to any type of field
operator -- bosonic or fermionic, relativistic or non-relativistic.
The resulting generalised form of (\tseref{G4}) then enables
(\tseref{N40}) to be generalised to all orders and to all types of
fields.  We also consider the symmetry properties of $n$-point normal
ordered products, and finally show how different splits may be related.

\subsection{The ``Trace Theorem''}

This is described in detail in
\tsecite{Ga, FW} and uses the cyclic property of the trace.

Consider again
\begin{equation}
{\cal{W}}_{1,2,\ldots, n} \; = \; \ll
\widehat{\psi}(x_1) \widehat{\psi}(x_2)
\ldots \widehat{\psi}(x_n) \gg,
\tseleq{thermn}
\end{equation}
and write
\begin{equation}
\widehat{\psi}(x) = \sum_{\bf k} \sum_{a=1}^{2} V_{\bf k} \eta({\bf k},
x) \hat{\alpha}_{a}({\bf k}).
\tseleq{trace1}
\end{equation}
Here $\hat{\alpha}_{1}({\bf k}) = \hat{a}_{\bf k}$ and
$\hat{\alpha}_{2}({\bf k}) = \hat{a}^{\dagger}_{\bf k}$ (we still
consider a real scalar relativistic field theory for
notational simplicity).  The thermal expectation value reduces to a
trace over
$\hat{\alpha}$ operators, which we call $S:= {\rm Tr} \left\{ e^{-\beta
\hat{H}_0} \hat{\alpha}_{a}({{\bf k}_1}) \hat{\alpha}_{b}({{\bf k}_2})
\ldots \hat{\alpha}_{t}({{\bf k}_n}) \right\}$. Since the commutation
relations of the $\hat{\alpha}$'s are known
$\hat{\alpha}_{a}({{\bf k}_1})$ can be commuted through
$\hat{\alpha}_{b}({{\bf k}_2})$, then $\hat{\alpha}_{c}({{\bf k}_3})$,
and so on until it is finally commuted through
$\hat{\alpha}_{t}({{\bf k}_n})$.  Then using the cyclic property of the
trace, it can be taken to the left-hand side of the expression.
Finally, after commuting through $e^{-\beta \hat{H}_0}$ one obtains
\begin{eqnarray}
\left( 1 - e^{\lambda_{a} \beta \omega_{{\bf k}_{1}}} \right) S  & = &
 {\rm Tr} \left\{ e^{-\beta
\hat{H}_0} \hat{\alpha}_{c}({{\bf k}_3}) \ldots \hat{\alpha}_{t}({{\bf
k}_n}) \right\}
\left[\hat{\alpha}_{a}({{\bf k}_1}), \hat{\alpha}_{b}({{\bf k}_2})
\right]
\nonumber
\\
 & + &  {\rm Tr} \left\{ e^{-\beta
\hat{H}_0} \hat{\alpha}_{b}({{\bf k}_2}) \ldots \hat{\alpha}_{t}({{\bf
k}_n}) \right\}
\left[\hat{\alpha}_{a}({{\bf k}_1}), \hat{\alpha}_{c}({{\bf k}_3})
\right]
\nonumber
\\
 & + &
\ldots
\nonumber
\\
 & + &  {\rm Tr} \left\{ e^{-\beta
\hat{H}_0} \hat{\alpha}_{b}({{\bf k}_2}) \ldots \hat{\alpha}_{t-1}({{\bf
k}_{n-1}}) \right\}
\left[\hat{\alpha}_{a}({{\bf k}_1}), \hat{\alpha}_{t}({{\bf k}_n})
\right] ,
\tselea{trace2}
\end{eqnarray}
where $\lambda_{1} = -1$, and $\lambda_{2} = 1$.
Now, one shows that
\begin{equation}
\frac{ \left[\hat{\alpha}_{a}({{\bf k}_1}), \hat{\alpha}_{b}({{\bf
k}_2}) \right] }{ \left( 1 - e^{\lambda_{a} \beta \omega_{{\bf k}_{1}}}
\right) } =
\delta_{a,1} \delta_{b,2} \ll \hat{a}_{{\bf k}_1}
\hat{a}^{\dagger}_{{\bf k}_2}
\gg  + \delta_{a,2} \delta_{b,1} \ll \hat{a}^{\dagger}_{{\bf k}_1}
\hat{a}_{{\bf k}_2} \gg.
\tseleq{contraction}
\end{equation}
Repeated use of (\tseref{trace2}) and
(\tseref{contraction}) enable ${\cal{W}}_{1,2,\ldots, n}$ to
be expressed as a sum of products of the form ${\cal{W}}_{1,2,\ldots,
n-2} {\cal{W}}_{n-1,n}$.  Iterating the argument, one obtains a
generalisation of (\tseref{G4}), that is
\begin{equation}
G^{2n}(x_1,\ldots,x_{2n}) = \sum_{{\rm perm}}
\left( \frac{1}{n!} \frac{1}{2} G_{{a_1},{a_2}} \ldots \frac{1}{2}
G_{{a_{2n-1}},{a_{2n}}} \right).
\tseleq{Greeny}
\end{equation}
The sum takes $a_1, \ldots a_{2n}$ through all
permutations of $1, \ldots ,2n$.
Clearly this argument has no reference to volume.

\subsection{The Path Integral}

Equation (\tseref{Greeny}) may be obtained by
using the path integral approach to thermal field theory.  Let $C$ be a
directed path  in the complex time plane whose end is $-i \beta$ below
the starting point, and let
$T_{C}$ denote path ordering of the operators with respect to $C$.  The
generating functional is then defined by
\begin{equation}
Z_{0}[j] := {\rm Tr} \left\{ e^{-\beta \hat{H}_0} T_{C}
\exp \left\{\int_{C} dt \int d^{3} {\bf x} \; j(t,{\bf x}) \phi(t,{\bf
x}) \right\} \right\},
\tseleq{generating1}
\end{equation}
where $j(t,{\bf x})$ are classical sources coupled to the
classical fields $\phi(t,{\bf x})$.  This can be rewritten as
\tsecite{LvW}
\begin{equation}
Z_{0}[j] =  \exp \left\{ - \frac{i}{2}
\int_{C} dt dt' \int d^{3} {\bf x}  d^{3} {\bf x'} \;j(t, {\bf x})
\Delta_{C}(x - x') j(t', {\bf x'})  \right\}.
\tseleq{generating2}
\end{equation}
Here $\Delta_{C}(x - x') = G_{1,2}$ is the two-point
thermal Green function or propagator -- it is the Green function of the
classical equation of motion for the fields $\phi(t,{\bf x})$ subject to
the KMS boundary condition \tsecite{LvW}.  As usual, the higher order
Green functions are given by
\begin{equation}
 G^{n}(x_1,\ldots,x_n) =  \left. \frac{\delta^{n} Z_{0}[j]}{\delta j(x_1)
\ldots j(x_n)} \right|_{j=0},
\tseleq{Greenfnl}
\end{equation}
and so (\tseref{Greeny}) is once again obtained.
Again there are clearly no volume corrections of the type mentioned in
the introduction.

\subsection{Expectation values of normal ordered products}

We now consider thermal expectation values of normal ordered products
of any type of field operators.  Since both the path-integral and ``trace
theorem'' can be generalised to {\em all} types of field operators
\tsecite{LvW,Ga}, (\tseref{Greeny}) can be generalised to give
\begin{equation}
 G^{2n}(x_1,\ldots,x_{2n}) = \sum_{{\rm perm}} (-1)^p
\left( \frac{1}{n!} \frac{1}{2} G_{{a_1},{a_2}} \ldots \frac{1}{2}
G_{{a_{2n-1}},{a_{2n}}} \right).
\tseleq{Green2}
\end{equation}
 However, from (\tseref{Wngen}) we know
the relation between time-ordered products, normal ordered products and
contractions.  Suppose we work with the splits satisfying
(\tseref{split1}) and (\tseref{split2}) where the expectation value of
the two-point normal ordered products is zero and the contraction is
equal to the thermal  propagator.  Then (\tseref{Green2}), with
(\tseref{W4}) for the four point case,   imply that
$\ll N_{1,2,3,4} \gg = 0$.  This generalises the result proved directly
for real relativistic scalar fields in section 4.

Now we combine these results for two- and four-point functions  with the
six-point versions of  (\tseref{Green2}) and
(\tseref{Wngen}); they imply that $\ll N_{1,2,3,4,5,6}\gg =
0$.   Repeating, we see that the splits satisfying (\tseref{split1}) and
(\tseref{split2}) ensure that the thermal expectation values of all
normal ordered functions are zero,
\begin{equation}
\ll N_{1,2,\ldots,2n}\gg = 0  .
\tseleq{NN0}
\end{equation}
All of these results are independent of volume.

\subsection{Symmetry of normal ordered products}

By definition the time-ordered function has the
symmetry property $T_{1,2,\ldots,n} = (-1)^p T_{a,b,\ldots,t}$ where
$a,b,\ldots,t$ is any permutation of $1,2,\ldots,n$ and $p$ is the
number of time pairs of fermionic fields are interchanged in moving from
an order $1,2,\ldots,n$  to an $a,b,\ldots,t$ order of fields.  Using
(\tseref{symmno}) it can be shown that
\begin{equation}
\left( D_{1,2} = \sigma D_{2,1} \Leftrightarrow   N_{1,2} = \sigma
N_{2,1} \right) \Rightarrow  N_{1,2,\ldots,n} = (-1)^p N_{a,b,\ldots,t} .
\tseleq{symmnogen}
\end{equation}
Thus for both the $T=0$ split and that of (\tseref{split1soln}) and
(\tseref{split2soln}), the normal ordered products have the same
symmetry as time ordered products.

\subsection{Relating two different splits}

Equation (\tseref{genWick}) may be used to derive a relationship
between different splits, i.e. between different types of normal
ordering. Suppose $N$ and $N'$ label two different kinds of normal
ordering (for which $D$ and $D'$ are the corresponding c-number
contractions).  Then from (\tseref{genWick}) they are related by
\begin{eqnarray}
\lefteqn{
N' \left[ \exp \{ -i \int d^{4} x \; j_i(x) \psi_i(x) \} \right]
 =
N \left[ \exp \{ -i \int d^{4} x \; j_i(x) \psi_i(x) \} \right] }
\nonumber
\\
& \times &
\exp \{ - \frac{1}{2} \int d^{4} x d^{4} y \;
 j_i(x) \left( D [ \psi_i(x) \psi_j(y) ]
- D'[ \psi_i(x) \psi_j(y) ] \right) j_j(y) \}.
\tselea{genWick2}
\end{eqnarray}
After functional differentiation this yields
\begin{eqnarray}
\lefteqn{
\sum_{\rm perm} \frac{(-1)^p}{(2n)!} {N'}_{a_1,a_2,\ldots,a_{2n}}
=
}
\nonumber
\\
& &
\sum_{\rm perm} {(-1)^p}
\left[ \sum_{m=0}^{n} \frac{1}{(2(n-m))!} {N}_{a_1,\ldots,a_{2(n-m)}}
. \frac{1}{2^m m!} \prod_{j=m+1}^{n} (D-D')_{a_{2j-1},a_{2j}} \right]
\tselea{difsplit}
\end{eqnarray}
The difference between normal ordering is
therefore related to the difference between contractions.  Now let $D'$
and $N'$ refer to the usual $T=0$ split.  Then
${D'}_{1,2}=G_{1,2}(T=0)$ and $\ll N' \gg \neq 0$.  Thus in
calculating the thermal expectation value of a $n$-point time ordered
function using (\tseref{Wn}), the terms containing $\ll N' \gg $ are
non-vanishing.  (Recall that for this $T=0$ split normal ordered
products are symmetric, and so the sum over permutations in
(\tseref{difsplit}) simplifies greatly.)

Now compare the $T=0$ split to the splits satisfying
$(\tseref{split1})$ and (\tseref{split2}).  Let these splits have
corresponding normal ordering and contraction labelled by $N$ and $D$.
Then $\ll N \gg = 0$, and taking the thermal expectation value of
(\tseref{difsplit}) gives
\begin{equation}
\ll {N'}_{1,2,\ldots,2n}\gg = \sum_{\rm perm} \frac{(-1)^p}{2^n n!}
\prod_{j=1}^n (D-D')_{a_{2j-1},a_{2j}}  = \sum_{\rm
perm} \frac{(-1)^p}{2^n n!} \prod_{j=1}^n (G(T)-G(T=0))_{a_{2j-1},a_{2j}}
\tseleq{T0no}
\end{equation}
Hence if one works with the usual $T=0$ split, the additional terms
coming from $\ll N' \gg \neq 0$ are accounted for precisely by the
thermal corrections $G(T)-G(T=0)$ to the zero temperature propagator.
It may therefore make more sense to
work with the splits  defined by $(\tseref{split1})$ and
(\tseref{split2}).

We can use the formula (\tseref{T0no}) to provide a different view of
several aspects of thermal field theory.  For instance in real-time
approaches there is a doubling of the degrees of freedom
\tsecite{LvW,vW,Raybook,Mi,TFD,SU,He}.  It is only the finite
temperature corrections to the zero temperature propagators which
mix the `physical' and `unphysical' degrees of freedom.  Thus one
could look on the doubling as being necessary only to take account
of the normal ordered products, $N'$, where they are defined in the
usual manner of zero temperature.  

\section{Conclusion}

Contrary to claims in the literature, we have proved that for fields of
arbitrary statistics there are no finite volume corrections  to the
usual relationship between thermal Green functions (\tseref{Green2}).
Thus the operator approaches and path integral approaches to thermal
field theory are indeed consistent in this respect.  So in trying to
understand the physics of the
quark-gluon plasma formed in relativistic heavy-ion collisions, we have
to look elsewhere to understand how finite volume affects the
physics.

In proving the above, we have defined and used a linear split of the
field operators into positive and negative parts such that $\ll
N_{1,2,\ldots,2n}\gg = 0$.  This split enables usual $T=0$ field
theory to be mimicked.  By choosing to work with a split where such
a result holds, we will simplify practical calculations.  

For instance when we are working with unphysical degrees of freedom
at finite temperature, such as appear in gauge theories, one usually
thermalises all the degrees of freedom.  However, it has been noted
that such degrees of freedom need not be thermalised \tsecite{LR}. 
In the language of this paper, we would say that one may choose a
different split (\tseref{decomp}) for each field.  In particular, we
may optimise the splits depending on how each field acts on the
states in the ensemble.  So with unphysical modes and an ensemble
average taken over just physical states, the best split, normal
ordering etc.\ for the unphysical modes is the zero temperature one,
whereas we normal order the physical modes in terms of the usual
thermal split of (\tseref{split1}) and (\tseref{split2}).  
This idea can extended to out-of-equilibrium situations where
different physical degrees of freedom may well be present in the
ensemble in quite different amounts suggesting that we define normal
ordering and contractions differently for each field so as to ensure
that $\ll N_{1,2,\ldots,2n}\gg = 0$ always.  

We have refered only to the path-ordered approaches to thermal field
theory, be they real- or imaginary-time formalisms.
Another simple method of obtaining (\tseref{Green2}) is to
use TFD \tsecite{LvW,vW,Raybook,TFD,SU}.
There one must normal order with respect to the ``thermal
quasi-particle operators'' ${\xi}, \widetilde{{\xi}}, {\xi}^{\sharp}$
and $ \widetilde{{\xi}}^{\sharp}$ which annihilate the thermal
vacuum, as this guarantees that all thermal expectation values of normal
ordered products vanish, for example see \tsecite{Re}.
(For definition of the terminology and
notation, see \tsecite{He}, \tsecite{SU}.)  Equation (\tseref{Green2})
then follows directly from (\tseref{Wn}) whereas in \tsecite{SU} a
rather lengthy derivation is given.  This argument is also
clearly independent of volume as it mimics the standard zero
temperature field theory case.  It is immediately obvious that these
comments regarding TFD and normal ordered products
apply to any situation where a Bogoliubov transformation is used.

\section*{Acknowledgements}

We thank Tom Kibble for many useful comments.  T.S.E. thanks  the
Royal Society for their support.  D.A.S. is supported by P.P.A.R.C.\
of the U.K.

\typeout{--- No new page for bibliography ---}

\end{document}